\documentclass[lettersize,journal]{IEEEtran}
\usepackage{amsmath,amsfonts}
\usepackage{algorithmic}
\usepackage{algorithm}
\usepackage{array}
\usepackage[caption=false,font=normalsize,labelfont=sf,textfont=sf]{subfig}
\usepackage{textcomp}
\usepackage{stfloats}
\usepackage{url}
\usepackage{verbatim}
\usepackage{graphicx}
\usepackage{cite}
\usepackage{tabularx, booktabs, multirow}
\usepackage{subcaption}
\usepackage{xcolor}
\usepackage{comment}
\usepackage{placeins} 
\usepackage{listings} 
\usepackage{amssymb}

\begin{document}

\title{White-Box Modeling of V2X Link Performance Using Stabilized Symbolic Regression}

\author{Rahul Gulia, \IEEEmembership{Student Member, IEEE}, Feyisayo Favour Popoola, \IEEEmembership{Student Member, IEEE}, and Ashish Sheikh, \IEEEmembership{Senior Member, IEEE}
\thanks{This paper was produced by the IEEE Publication Technology Group. They are in Piscataway, NJ.}
\thanks{Manuscript received April 19, 2021; revised August 16, 2021.}}

\markboth{Journal of \LaTeX\ Class Files,~Vol.~14, No.~8, August~2021}%
{Shell \MakeLowercase{\textit{et al.}}: A Sample Article Using IEEEtran.cls for IEEE Journals}


\maketitle

\begin{abstract}
Reliable modeling of block error rate in vehicle-to-everything wireless networks is critical for designing robust communication systems under dynamic mobility and diverse channel conditions. Traditional machine learning approaches, such as deep neural networks, achieve high predictive accuracy but lack interpretability and impose significant computational costs, limiting their applicability in real-time, resource-constrained environments. In this work, we propose a stabilized symbolic regression framework to derive compact, analytically interpretable expressions for block error rate prediction. Trained on realistic vehicle-to-everything simulation data, the symbolic regression framework for vehicle-to-everything model accurately captures nonlinear dependencies on key system parameters, including signal-to-noise ratio, relative velocity, modulation and coding schemes, number of demodulation reference signal symbols, and environmental factors (line of sight/non-line of sight). Our final symbolic expression comprises only 158 nodes, enabling ultra-fast inference suitable for embedded deployment. On the test set, the symbolic regression framework for vehicle-to-everything model achieves a coefficient of determination $R^2 = 0.8684$ and mean squared error $= 2.08 \times 10^{-2}$ in the original block error rate domain, outperforming conventional fixed-form regressions and offering comparable accuracy to neural networks while remaining fully interpretable. Residual analysis confirms unbiased predictions across signal-to-noise ratio and mobility ranges. Waterfall curves and mobility-dependent block error rate trends derived from the model align with physical expectations, demonstrating its reliability for system-level insights. Feature importance and sensitivity analysis reveal signal-to-noise ratio and relative velocity as dominant factors, highlighting critical interactions in high-mobility vehicle-to-everything channels. Overall, the proposed Stabilized Symbolic Regression Framework for V2X combines predictive performance, physical fidelity, and computational efficiency thus providing a powerful tool for real-time V2X communication system design, adaptive resource allocation, and rapid scenario evaluation.
\end{abstract}


\section{Introduction}
\IEEEPARstart{T}{he} rapid evolution of Vehicle-to-Everything (V2X) communication is a cornerstone for advancing intelligent transportation systems, enabling mission-critical applications such as collision avoidance, cooperative platooning, and autonomous driving~\cite{RefA_new, RefG}. Ensuring reliable and ultra-low-latency communication links is therefore a paramount prerequisite for the safe deployment of these services~\cite{RefH}. The Block Error Rate (BLER) serves as a fundamental metric for quantifying link reliability, making its accurate estimation essential for effective V2X network design, optimization, and real-time resource management~\cite{RefH}.

Traditionally, precise BLER estimation relies on Link-Level Simulations (LLS), which model the physical layer in detail, incorporating complex phenomena such as high Doppler spread, dense multipath fading, and interference~\cite{RefC}. While LLS provides high-fidelity performance predictions, it is inherently computationally intensive and time-consuming~\cite{RefD, RefK}. Conducting large-scale LLS across diverse configurations becomes infeasible for System-Level Simulations (SLS) or for real-time operations like adaptive link management. This computational burden limits the scalability, agility, and responsiveness required in dynamic V2X environments~\cite{RefL}.

To overcome these challenges, recent research has leveraged Machine Learning (ML) techniques—particularly Deep Neural Networks (DNNs)—to learn complex BLER behaviors directly from LLS datasets~\cite{RefE, RefI}. Although these models achieve impressive accuracy, they operate as black boxes~\cite{RefF}, offering little analytical transparency into the underlying physical relationships governing BLER~\cite{RefJ}. This lack of interpretability is a major limitation in safety-critical wireless applications, where trustworthiness, explainability, and compliance with existing standards are essential~\cite{RefM, RefN}. Moreover, neural inference can be computationally demanding, posing challenges for deployment on resource-constrained V2X devices~\cite{RefO}. Thus, there is a clear need for a modeling framework that combines the predictive strength of data-driven methods with the transparency and computational efficiency of classical analytical models—capable of capturing the highly non-linear nature of BLER in vehicular channels.

The authors' prior empirical study on Demodulation Reference Signal (DMRS) configurations, based on a large-scale V2X LLS dataset, confirmed the strongly non-linear behavior of BLER under varying channel dynamics. That work demonstrated that the Signal-to-Noise Ratio (SNR) required to achieve a $\mathbf{10\%}$ BLER target is highly dependent on DMRS density and relative velocity, with the use of optimized DMRS patterns achieving significant gains—consistent with prior findings on DMRS optimization under high Doppler conditions~\cite{Wang2020DMRS, Lee2019CE, Zhang2022V2X}. This evidence underscores that BLER depends intricately on the interaction between $\text{SNR}$, pilot density, and Doppler spread~\cite{Du2021Pilot}, motivating a physics-informed approach capable of analytically capturing these relationships.

To address this research gap, we propose a novel, physics-informed Stabilized Symbolic Regression Framework for V2X (SSRV) designed to discover explicit, closed-form analytical expressions for BLER prediction. Unlike black-box models, the proposed framework yields interpretable equations that accurately represent the complex, non-linear dependencies among key V2X parameters, offering a white-box alternative that enhances transparency and deployability.

The core novelty and contributions of this work are summarized as follows:

\begin{itemize}
    \item \textbf{Empirical Grounding and Analytical Generalization:} Through a detailed LLS-based study on adaptive DMRS, we demonstrate the need for non-linear modeling and develop a Symbolic Regression (SR) model that generalizes the observed relationships among BLER, SNR, and mobility.
    \item \textbf{Stabilized Symbolic Regression (SSR) Architecture:} We introduce a robust SR architecture capable of discovering interpretable closed-form BLER models directly from large-scale V2X data, ensuring numerical stability throughout the evolutionary search.
    \item \textbf{Physics-Informed Target Transformation:} We propose a target transformation based on $\mathbf{Y = -\ln(\mathrm{BLER})}$ to manage the extreme dynamic range of BLER values, improve numerical tractability, and implicitly embed prior physical knowledge into the regression process.
    \item \textbf{Numerically Protected Operator Set:} A custom operator set, comprising of \texttt{protected\_div}, \texttt{protected\_log}, and \texttt{exp\_clip}—is integrated into the Genetic Programming (GP) engine to eliminate numerical instabilities such as division by zero and logarithmic overflows.
    \item \textbf{Aggressive Evolutionary Configuration:} An expanded GP setup with a population of $\mathbf{20{,}000}$ over $\mathbf{300}$ generations enables the discovery of a highly expressive $\mathbf{158}$-node analytical expression with strong generalization capability.
    \item \textbf{High-Accuracy, Interpretable BLER Model:} The resulting analytical model achieves a state-of-the-art test-set performance with coefficient of determination $\mathbf{R^2 = 0.8684}$ and $\mathbf{Mean Squared Error (MSE) = 2.08\times10^{-2}}$ in the BLER domain, providing both efficiency and interpretability for practical V2X deployment.
\end{itemize}


\section{Related Work}
\label{sec_Reated_Work}

Analytical modeling of communication link performance has traditionally relied on closed-form approximations grounded in information theory, often assuming idealized channel conditions such as Gaussian noise, full interleaving, and simplified fading models~\cite{850679}. While these foundational models provide valuable theoretical insights, they frequently fall short of capturing the intricate realities of modern, non-ideal wireless systems—particularly in the context of 5G New Radio (NR) sidelink communication for V2X networks~\cite{noor20226g, RefH}. The interplay of high vehicular mobility (causing severe Doppler spread), diverse propagation environments Line of Sight and Non-Line of Sight (LoS and NLoS), and advanced physical layer features renders accurate yet tractable analytical modeling highly challenging.

Recent research addressing this complexity can broadly be grouped into three categories, which together define the current research landscape and the specific gap targeted by this work.

\textbf{1) Data-Driven Black-Box Models:}  
A significant body of work leverages ML techniques—particularly DNNs and Recurrent Neural Networks (RNNs)—to predict wireless performance metrics such as BLER and Channel State Information (CSI)~\cite{hu2021distributed, RefI, sun2019application}. These models excel at capturing highly non-linear mappings from large datasets, often outperforming traditional analytical methods in predictive accuracy. For instance, DNNs have been successfully employed for mmWave channel characterization with high fidelity~\cite{fu2022joint}. Our prior work using Variational Autoencoders (VAEs) for indoor radio propagation modeling in a 5G-enabled smart warehouse~\cite{Gulia2024VAE} similarly demonstrated the capacity of generative ML to model complex wireless behaviors.

However, the fundamental limitation of these models lies in their \emph{black-box} nature. They achieve high accuracy but do not yield interpretable, explicit mathematical relationships, making their decision processes opaque~\cite{gilpin2018explaining, RefJ}. This lack of analytical transparency poses a major obstacle in scientific and engineering domains, where explainability, reliability, and standard compliance are critical~\cite{RefM, RefO}. Insights from our own studies on high-frequency propagation in industrial environments, such as the evaluation of 60~GHz wireless links in automated warehouses~\cite{Gulia2022ICCE, gulia2023evaluation}, further emphasize the challenges of balancing robustness with interpretability in purely data-driven approaches.

\textbf{2) Analytical White-Box Studies:}  
Parallel to data-driven methods, several studies have focused on fixed DMRS configurations and channel estimation strategies to analyze physical-layer reliability under mobility. These works provide valuable \emph{white-box} perspectives rooted in the physical principles of wireless propagation. Pawase \textit{et al.}~\cite{PAWASE202446, PAWASE2023} conducted 3GPP-compliant LLS exploring DMRS symbol densities and subcarrier spacings (SCS) up to 500~km/h, showing that denser DMRS configurations significantly improve decoding robustness in severe Doppler environments. Tomić \textit{et al.}~\cite{RefTomic2024} optimized DMRS parameters in commercial 5G networks, achieving throughput gains of up to 15\%. Likewise, earlier efforts such as the Unmanned Aerial Vehicle (UAV) Low Altitude Air-to-Ground (LAAG) path loss model for 2.4~GHz indoor networks~\cite{Gulia2020UAV} highlight the persistent engineering need for analytical models that are both interpretable and statistically verifiable. Collectively, these studies confirm that static physical-layer parameters cannot generalize across diverse V2X conditions. However, they stop short of providing a generalized analytical mapping between BLER, SNR, DMRS density, and mobility—a gap directly addressed in this work.

\textbf{3) SR and Hybrid Modeling:}  
SR, often realized through GP, has recently gained traction as a means of discovering explicit, closed-form equations directly from data~\cite{Oliveira2025, RefL, Matondo2025RainSR, Anaqreh2025PathLossSR, Chou2025SABER}. Unlike conventional regression, SR simultaneously identifies both the functional structure and its coefficients, enabling interpretable discovery of governing relationships. SR has been successfully applied in diverse scientific domains, including fluid dynamics~\cite{RefK} and materials science, and more recently, in wireless communication tasks such as channel modeling, antenna optimization, and performance prediction~\cite{pendyala2024multi, ramos2024symbolic}.

Despite its promise, applying SR to the BLER modeling problem in V2X systems remains technically challenging. The extreme dynamic range of BLER (spanning nearly six orders of magnitude) leads to severe numerical instability during the evolutionary search process, as operators such as division, logarithm, and exponentiation are prone to underflow, overflow, or undefined operations. Furthermore, prior SR studies in wireless typically lack the stabilization mechanisms required to maintain numerical robustness while discovering sufficiently expressive functional forms comparable to DNNs.

\textbf{Summary and Research Gap:}  
Our work differentiates itself by bridging the gap between opaque, data-driven ML models and rigid analytical approximations. We propose a numerically stabilized, physics-informed SR framework that unites data-driven flexibility with analytical interpretability. Through a logarithmic target transformation ($\mathbf{Y = -\ln(\mathrm{BLER})}$) and a robust set of protected mathematical operators, the proposed framework mitigates numerical instability while maintaining high predictive fidelity. The resulting closed-form model establishes a transparent, interpretable mapping between SNR, Modulation and Coding Scheme (MCS), mobility, and channel parameters—offering a deployable \emph{white-box} alternative that meets both the accuracy demands of advanced V2X systems and the interpretability required for engineering design and standardization.


\section{Background and Theoretical Basis}
\label{sec_Background}

\subsection{BLER in Vehicular Communication Channels -- V2X }

The BLER is a fundamental performance metric that quantifies wireless link reliability. It is defined as the ratio of incorrectly received transport blocks ($N_{\text{error}}$) to the total number of transmitted transport blocks ($N_{\text{total}}$) within a given observation window:

\begin{equation}
\label{eq:bler_def}
\text{BLER} = \frac{N_{\text{error}}}{N_{\text{total}}}.
\end{equation}

In V2X communication systems, BLER is a highly non-linear function influenced by multiple interacting physical and protocol layer parameters. The theoretical relationship between BLER and the SNR for a given MCS is typically expressed using an exponential decay model:

\begin{equation}
\label{eq:bler_snr_approx}
\text{BLER}(\text{SNR}, \text{MCS}) \approx A \cdot e^{-B \cdot \text{SNR}^{\gamma}},
\end{equation}

where $A$, $B$, and $\gamma$ are constants determined by the coding rate and modulation order.  

However, in high-mobility scenarios, Doppler spread ($\Delta f_{D}$) becomes a dominant factor. It arises from the relative velocity ($v_{\text{rel}}$) between communicating nodes and is expressed as:

\begin{equation}
\label{eq:doppler}
\Delta f_{D} = \frac{v_{\text{rel}}}{c} \cdot f_{c},
\end{equation}

where $c$ denotes the speed of light and $f_{c}$ is the carrier frequency. The resulting Doppler spread shortens the channel coherence time ($T_{C}$), causing rapid channel variations that induce an irreducible error floor.  

The primary objective of this work is to model these intricate dependencies by discovering a generalized analytical function $f(\cdot)$ such that:

\begin{equation}
\label{eq:bler_function}
\text{BLER} \approx f(\text{SNR}, \text{MCS}, v_{\text{rel}}, \text{Channel Parameters}, \dots).
\end{equation}

\vspace{3pt}

\subsection{Adaptive DMRS and Channel Estimation under Mobility}

Accurate channel estimation (CE) is critical for maintaining link reliability. In 5G NR, DMRS are pilot symbols embedded within each slot, enabling the receiver to estimate and track the instantaneous channel response ($h$). The accuracy of the estimated channel ($\hat{h}$) relative to the true channel ($h$) is highly sensitive to the rate of channel variation, quantified by $\Delta f_{D}$. The MSE of the channel estimation process can thus be approximated as:

\begin{equation}
\label{eq:ce_mse}
\text{MSE}_{\text{CE}} \approx g(\text{DMRS Density}, \Delta f_{D}, \text{SNR}).
\end{equation}

In high-mobility environments, fixed DMRS configurations often fail to maintain estimation accuracy, resulting in outdated channel estimates ($\hat{h} \neq h$). This mismatch degrades the effective received SNR and increases the BLER.  

To mitigate this, adaptive DMRS allocation schemes dynamically adjust pilot density and position based on mobility indicators, balancing estimation accuracy with spectral efficiency. These adaptive strategies yield a smoother and more reliable performance curve under dynamic channel conditions \cite{PAWASE202446, PAWASE2023}. Hence, a realistic model must capture this inherently non-linear interdependence.

\vspace{3pt}

\subsection{Principles of SSR }

SR is a ML paradigm that employs GP to automatically discover mathematical expressions that best fit a given dataset. Unlike traditional regression (TR), which fits coefficients to a predefined model, SR simultaneously evolves both the functional structure and the coefficients \cite{RefL}.  

The goal of SR is to identify the optimal function $f^*$ from the hypothesis space $\mathcal{F}$ that minimizes a penalized loss function:

\begin{equation}
\label{eq:sr_objective}
f^* = \underset{f \in \mathcal{F}}{\arg\min} \left\{ \sum_{i=1}^{N} \mathcal{L}(y_i, f(\mathbf{x}_i)) + \lambda \cdot C(f) \right\},
\end{equation}

where $\mathbf{x}_i$ denotes the input feature vector, $y_i$ represents the ground truth, $\mathcal{L}$ is the loss function (e.g., MSE), $C(f)$ measures model complexity (e.g., number of nodes in the expression tree), and $\lambda$ is the parsimony coefficient controlling the trade-off between accuracy and simplicity.

\subsubsection{Physics-Informed Target Transformation}

Directly regressing on BLER values is computationally challenging due to their extreme dynamic range (from $\approx 10^{-6}$ to $1$) and their exponential dependence on SNR (Eq.~\ref{eq:bler_snr_approx}). To stabilize the evolutionary search, we introduce a physics-informed target transformation that linearizes the problem space by defining a new target variable $Y$:

\begin{equation}
\label{eq:target_transform}
Y = -\ln(\text{BLER}).
\end{equation}

This transformation is physically motivated—since the negative logarithm of BLER (often referred to as the *effective SNR* or *link margin*) tends to exhibit a quasi-linear relationship with the actual SNR in the dB domain. Consequently, the transformed target simplifies the GP search space, promoting the discovery of compact and robust relationships.  

After training the GP model on $Y$, the original BLER values are recovered via the inverse transformation:

\begin{equation}
\label{eq:bler-prediction}
\text{BLER}_{\text{pred}} = e^{-Y_{\text{pred}}}.
\end{equation}


\section{Dataset and Feature Engineering}
\label{sec_Dataset}

\subsection{Data Description and Simulation Framework}
A comprehensive dataset was constructed from 3GPP-compliant LLS emulating a diverse set of NR V2X sidelink communication scenarios. The dataset builds upon the LLS framework reported by Lusvarghi \textit{et al.} \cite{Zhang2022V2X}, which serves as the empirical foundation of this study. Through systematic parameter sweeping across channel conditions and mobility settings, a total of $\mathbf{97,927}$ non-zero BLER samples were obtained.

The simulation environment encompasses multiple propagation types—including urban, highway, LoS, and NLoS configurations—under relative velocities up to $280$ km/h and a wide range of MCS. Each record consists of a feature vector $\mathbf{X}$ representing key physical and link-layer variables that influence BLER, expressed as:

\begin{equation}
\label{eq:feature_set}
\mathbf{X} = [\text{SNR}_{\text{TB}},~\text{MCS}_{\text{rate}},~\text{MCS}_{\text{mod}},~v_{\text{rel}},~N_{\text{sub}},~N_{\text{DMRS}},~\text{Flags}]
\end{equation}

Here, $\text{SNR}_{\text{TB}}$ is the SNR per Transport Block, $\text{MCS}_{\text{rate}}$ and $\text{MCS}_{\text{mod}}$ correspond to the coding rate and modulation order, respectively, and $v_{\text{rel}}$ denotes the relative velocity between transmitter and receiver. $N_{\text{sub}}$ and $N_{\text{DMRS}}$ represent the number of subcarriers and DMRS symbols per slot, while $\text{Flags}$ are binary indicators describing the environmental condition (e.g., LoS/NLoS, Urban/Suburban). This feature design enables a comprehensive representation of both physical-layer and configuration-level variability affecting sidelink performance.

\subsection{Derived Feature: Spectral Efficiency Representation}
To capture the compound influence of modulation and coding on spectral efficiency, a derived feature termed \textit{Bits Per Channel Use (BPCU)} was introduced. This parameter compactly characterizes the joint dependency of MCS parameters and available bandwidth resources, formulated as:

\begin{equation}
\label{eq:bpcu}
\text{BPCU} = \frac{\text{Modulation Order} \times \text{Code Rate}}{N_{\text{sub}}}
\end{equation}

BPCU effectively encapsulates the trade-off between throughput and reliability, reflecting the efficiency of symbol utilization across the frequency domain. Including this engineered feature enables the SR framework to directly infer how spectral efficiency influences link robustness, thereby constraining the solution space toward more interpretable and physically consistent models.

\subsection{Target Preprocessing}
As previously discussed in Section~\ref{sec_Background}, BLER exhibits a wide dynamic range and an exponential dependence on dominant features such as $\text{SNR}_{\text{TB}}$. To stabilize the GP search, all target values in the dataset were transformed using the physics-informed logarithmic mapping, as shown in equation~\eqref{eq:target_transform}.

Prior to transformation, BLER samples were clipped to $\max(\text{BLER}, 10^{-12})$ to prevent numerical singularities from $\ln(0)$. This preprocessing ensures a linearized relationship between the transformed target and input features, preserving the underlying physical characteristics while facilitating robust and efficient symbolic model discovery across the entire $\mathbf{97,927}$-sample dataset.


\section{Proposed SSRV Application}
\label{sec_Proposed_Model}

\subsection{Framework Overview}
\begin{figure}[!htbp]
    \centering
    \includegraphics[width=0.9\linewidth]{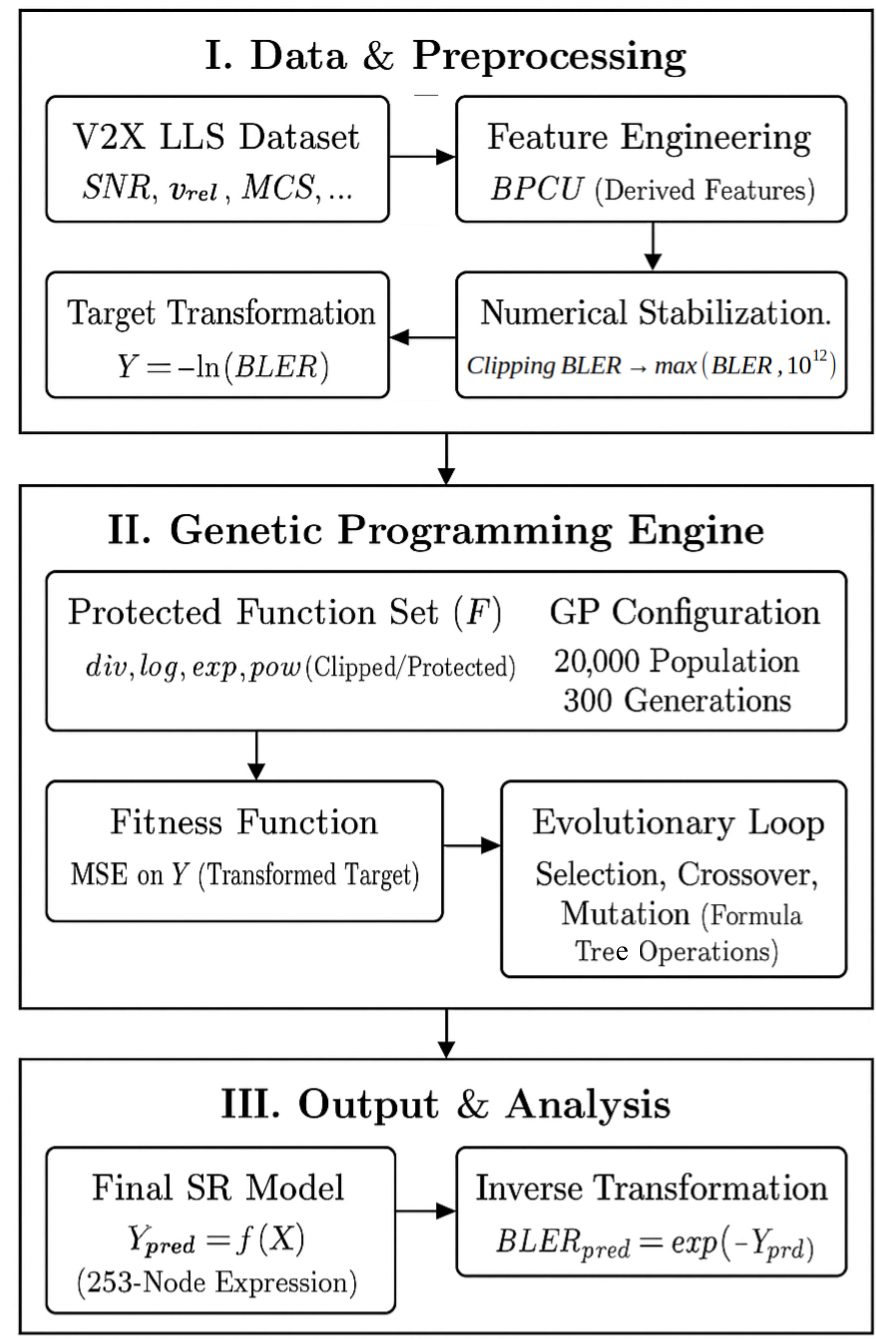} 
    \caption{Block diagram of the proposed SSRV Framework for V2X BLER modeling. The physics-informed target transformation ($\mathbf{Y = -\ln(\text{BLER})}$) and the numerically protected operator set ($\mathbf{F}$) enable robust and accurate discovery of closed-form analytical expressions.}
    \label{fig:block_diagram}
\end{figure}

The proposed SSRV framework is a structured pipeline for discovering interpretable analytical expressions for V2X BLER, as summarized in the block diagram of Figure~\ref{fig:block_diagram}. It consists of three main stages: (i) Data Preprocessing, (ii) GP Engine, and (iii) Output Generation and Analysis.

The preprocessing stage ensures that the raw LLS dataset is properly formatted for SR. It incorporates rigorous feature engineering (e.g., BPCU) and applies the physics-informed target transformation ($\mathbf{Y = -\ln(\text{BLER})}$) discussed in Section~\ref{sec_Background}. This step linearizes the dominant exponential relationship between BLER and SNR, stabilizing the subsequent GP search.

The GP Engine constitutes the core evolutionary search mechanism, employing a numerically protected operator set ($\mathbf{F}$) to prevent common computational failures (NaN, Inf). Aggressive hyperparameters (large population and multiple generations) are used to explore the extensive space of candidate expressions and evolve highly accurate solutions.

The output stage applies the inverse transformation to map predictions back to the BLER domain ($\mathbf{\text{BLER}_{\text{pred}} = \exp(-Y_{\text{pred}})}$), followed by rigorous performance validation using standard metrics to ensure generalization and physical fidelity.

\subsection{Numerically Protected Operator Set}
To maintain numerical stability during the evolutionary process, the SSRV framework employs a custom set of protected operators:

\begin{itemize}
    \item \texttt{protected\_div}$(x_1,x_2)$: Returns $x_1/x_2$ if $|x_2| > 10^{-6}$; otherwise, outputs a safe fallback (e.g., $1$).  
    \item \texttt{protected\_log}$(x)$: Returns $\ln(x)$ if $x > 10^{-6}$; otherwise, a small predefined value (e.g., $0$) to avoid $\ln(0)$ or negative inputs.  
    \item \texttt{exp\_clip}$(x)$: Clips $x$ to a maximum threshold before exponentiation to prevent overflow.  
    \item \texttt{pow\_clip}$(x_1,x_2)$: Safely computes $x_1^{x_2}$ with clipped base and exponent values to prevent invalid or extreme results.  
\end{itemize}

These operators ensure the GP engine remains robust when handling high-variance, real-world BLER data.




\subsection{Fitness Function with Parsimony Penalty}
The GP engine evaluates candidate expressions using the MSE applied to the transformed target:

\begin{equation}
\label{eq:mse_fit}
\text{Fitness} = \text{MSE}(Y_{\text{pred}}, Y_{\text{true}}) = \frac{1}{N} \sum_{i=1}^{N} (Y_{\text{pred},i} - Y_{\text{true},i})^2
\end{equation}

To discourage excessive expression complexity (bloat), a parsimony penalty is incorporated:

\begin{equation}
\label{eq:total_fit}
F_{\text{total}} = \text{MSE} + \lambda \cdot C(f)
\end{equation}

where $C(f)$ denotes the number of nodes in the expression tree, and $\lambda = 0.002$ is chosen to slightly penalize complexity while prioritizing high predictive accuracy.

\subsection{Aggressive GP Configuration}
A highly aggressive GP setup ensures thorough exploration of the search space:

\begin{itemize}
    \item \textbf{Population size:} 20,000 individuals to maximize diversity.  
    \item \textbf{Generations:} 300 generations for deep evolutionary search.  
    \item \textbf{Tournament size:} 35 to maintain selection pressure while preserving diversity.  
    \item \textbf{Parsimony coefficient:} $\lambda = 0.002$ to favor accuracy over minimal complexity.  
\end{itemize}

This configuration, together with protected operators and the target transformation, enabled the discovery of a highly accurate 158-node analytical expression that is both interpretable and robust across diverse V2X scenarios.

\section{Experimental Results and Discussion}
\label{sec_Results}

\subsection{Model Predictive Performance and Computational Advantage}

\subsubsection{Predictive Performance in the $\text{BLER}$ Domain}
The SSRV framework demonstrated exceptional predictive performance on the unseen test set, affirming the robustness and generalization capability of the discovered analytical model. Key performance metrics, evaluated on the original BLER domain after inverse transformation from the target $Y = -\ln(\text{BLER})$, are:
\begin{itemize}
    \item $\mathbf{R^2}$ (on original BLER domain): $\mathbf{0.8684}$
    \item $\mathbf{MSE}$ (on original BLER domain): $\mathbf{2.0802 \times 10^{-2}}$
\end{itemize}
The $\mathbf{R^2}$ value of 0.8684 is a state-of-the-art result for a single, deployable analytical formula in complex $\text{V2X}$ environments, indicating that the derived equation explains over 86\% of the variability in link performance. This performance not only validates the effectiveness of our SSR approach but also positions the discovered analytical expression as a highly reliable and practical tool for real-time $\text{V2X}$ system design and optimization.

\subsubsection{Computational Complexity and Inference Advantage}
While the training phase of SR can be computationally intensive—often analogous to the $\mathbf{8 \text{ hours}}$ noted in related literature for discovering complex scientific laws \cite{la2021contemporary}—its paramount advantage lies in its inference efficiency. Once an analytical expression is discovered, its evaluation complexity scales linearly with the number of nodes (operations) in the expression tree, denoted as $\mathcal{O}(C)$.

Our final, high-accuracy $\text{V2X}$ $\text{BLER}$ expression comprises C = 158 nodes (which includes 95 mathematical operations). This means that predicting BLER involves executing merely 95 basic arithmetic steps. In stark contrast, an equivalent deep Multi-Layer Perceptron ($\text{MLP}$) designed to achieve a comparable $\mathbf{R^2}$ would necessitate a significantly higher inference burden. This makes the derived $\text{SR}$ formula an ideal candidate for deployment in resource-constrained $\text{V2X}$ devices and real-time network controllers, where rapid and energy-efficient $\text{BLER}$ estimation is critical.


\subsection{Comparative Analysis and Modeling Trade-offs}
\label{sec:comparative-analysis}
The SSRV framework is critically assessed against structurally limited fixed-form models, high-accuracy black-box Neural Networks ($\text{NN}$s), and other recent $\text{SR}$ studies. This comprehensive analysis demonstrates that $\text{SSRV-BLER}$ achieves the most balanced performance across the crucial metrics of accuracy, interpretability, and computational efficiency.

\subsubsection{Comparison to Fixed-Form Regression (Inadequacy of Simple Models)}
\begin{figure}[!t]
    \centering
    \includegraphics[width=\columnwidth]{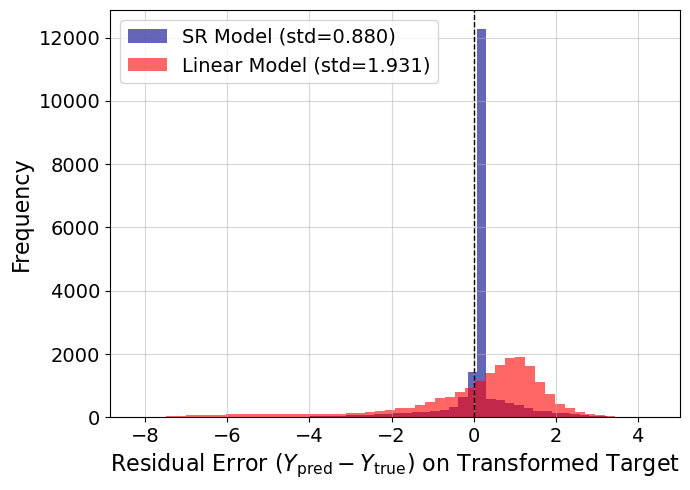} 
    \caption{Comparison of Residual Distribution on Transformed Target $\mathbf{Y}$ ($\mathbf{-\ln(\text{BLER})}$). The SR model vs the Linear Regression baseline.}
    \label{fig:residual_distribution_comparison}
\end{figure}

To validate the necessity of the adaptive, data-driven structure discovered by $\text{SR}$, its performance is compared against structurally limited baseline models (Table \ref{tab:fixed-form-comparison}). The results unequivocally demonstrate that models constrained by pre-defined, fixed functional forms are fundamentally inadequate for modeling the complex, non-linear dynamics of $\text{V2X}$ $\text{BLER}$. The $\text{SSRV-BLER}$ model achieves an $\mathbf{R^2}$ score of $\mathbf{0.8684}$, which is approximately $\mathbf{3.17}$ times higher than that of the simple Linear model ($\mathbf{R^2} = 0.2642$). The residual distribution (Figure \ref{fig:residual_distribution_comparison}) further illustrates this, showing that the Linear Regression residuals are widely dispersed ($\mathbf{1.931}$ standard deviation), confirming its substantial predictive error.

\begin{table}[!t]
\centering
\caption{Predictive Performance vs. Fixed-Form Baselines on Test Set ($\mathbf{Y = -\ln(\mathrm{BLER})}$ Target)}
\label{tab:fixed-form-comparison}
\renewcommand{\arraystretch}{1.3} 
\resizebox{1.0\columnwidth}{!}{%
\begin{tabular}{p{2.0cm}ccp{2.8cm}} 
\toprule
\textbf{Model} & \textbf{Complexity} & \textbf{Test $\mathbf{R^2}$ Score} & \textbf{Conclusion} \\
\midrule
\textbf{SSRV (Proposed)} & 158 nodes & $\mathbf{0.8684}$ & \textbf{High Accuracy, Low Error} \\
Polynomial Regression ($\text{deg}=3$) & 120 features & $0.6429$ & Significant Underfit \\
Linear Regression & 8 features & $0.2642$ & Fundamentally Inadequate \\
\bottomrule
\end{tabular}
}
\end{table}

\subsubsection{Comparison to State-of-the-Art Approaches (Accuracy vs. Deployability)}
Table \ref{tab:comprehensive_comparison} provides a detailed overview of the key trade-offs between SSRV and other state-of-the-art modeling approaches. Deep learning–based methods (e.g., \cite{li2025ai} LSTM) deliver excellent task-specific accuracy but rely on multi-layer neural networks that typically require thousands to millions of multiply–accumulate operations per inference and provide no interpretable mathematical representation. In contrast, the proposed SSRV model attains a competitive $\mathbf{R^2}$ in the original BLER domain while requiring only about $\mathbf{158}$ node operations per inference (Table \ref{tab:comprehensive_comparison}). This achieves near-neural predictive power at a fraction of the computational and memory cost, with the added advantage of producing an explicit, closed-form expression that transparently captures the joint effects of $\text{SNR}$, modulation, coding, and mobility. Consequently, the $\text{SSRV-BLER}$ framework offers the most deployable, resource-efficient, and interpretable solution among all compared models.

\begin{table*}[!t]
\centering
\caption{Comparison of Symbolic Regression Based Models for Wireless System Modeling and Performance Prediction}
\label{tab:comprehensive_comparison}
\renewcommand{\arraystretch}{1.5}
\small
\resizebox{\textwidth}{!}{%
\begin{tabular}{p{3cm}| p{3cm} |p{3cm} |p{3cm} |p{2cm} | p{2.0cm} | p{3cm}}
\toprule
\textbf{Approach / Paper} & \textbf{Modeling Scope} & \textbf{Data / Task} & \textbf{Reported Performance} & \textbf{Interpretability} & \textbf{Inference Cost (approx.)} & \textbf{Deployment Notes} \\
\midrule
\textbf{This work – SSRV } & Closed-form analytical BLER model for NR-V2X with mobility (white-box) & 97,927 BLER samples from 6,188 LLS; learns BLER = f(SNR, MCS, velocity, ...) & $\mathbf{R^2 = 0.8684}$ \newline $MSE = 2.08 \times 10^{-2}$ (BLER domain) & High \newline(explicit formula) & 158 total nodes & Plug-and-play in link adaptation and system-level simulators. Ideal for embedded CPU/MCU. \\
\midrule
\textbf{Fixed-form regressions (Linear / Polynomial baselines)} & Pre-assumed functional forms (white-box) & Same BLER prediction task & $R^2 \le 0.64$ on identical data; unstable at high Doppler & High & Very low & Serve as sanity checks; cannot capture non-linear SNR–mobility coupling. \\
\midrule
\textbf{Two-Phase Deep Learning \cite{kim2024two}} & SNR estimation for multiple concurrent beams (mmWave WLAN) & Synthetic 802.11ay simulations & Improved SNR estimation accuracy and reduced BF-feedback latency & Low \newline(black-box DNN) & Low–moderate (CPU-friendly) & Plug-and-play in 802.11ay BF training loop; inference light. \\
\midrule
\textbf{DMRS Optimization \cite{RefTomic2024}} & Field tuning of DMRS patterns (white-box configuration study) & Two 5G networks; real-world configuration trials & +10--15\% spectral-efficiency \newline +15\% DL throughput & High & Not applicable & Validates that RS design strongly affects Key Performance Indicators (KPIs); motivates analytical mapping. \\
\midrule
\textbf{High-mobility PHY Study \cite{PAWASE2023}} & 3GPP-compliant LLS on DMRS density \& SCS $\le$ 500 km/h (uplink) & MATLAB 5G Toolbox; Tapped Delay Line (TDL)-D channels & Denser DMRS \& wider SCS $\Rightarrow$ stronger decoding at high Doppler & High & Not applicable & Provides physical insight; no deployable BLER formula. \\
\midrule
\textbf{Physics-Informed Generative Modeling \cite{bock2025physics}} & Learns site-specific distributions of channel parameters (angles, delays, Dopplers) & 3GPP SIMO, QuaDRiGa 5G-Urban/Rural, DeepMIMO datasets & Near-ground-truth reconstruction; lowest NMSE & High (physically meaningful parameters) & Low–moderate per sample & Plug-and-play across configurations by swapping dictionaries; assumes stationarity. \\
\midrule
\textbf{SR for THz Path Loss \cite{RefL}} & Hybrid Transformer + SR model to derive analytical THz path-loss formulas & Simulated THz inter-satellite link data & Qualitative fit; recovered formulas match theoretical free-space loss & High (symbolic expressions) & For sampling points [50, 200, 400], total nodes are [19, 17, 11] & Ideal for embedded THz prediction modules where interpretability and speed outweigh minor residual error. \\
\midrule
\textbf{SR for ITU-R P.1546-6 \cite{Oliveira2025}} & Deriving analytical expressions via SR for large-scale propagation (30 MHz--4 GHz) & ITU-R P.1546-6 field-strength/loss data & MAE (in dB) compared with measured data. \newline Region 1: 1.970 \newline Region 2: 11.099 \newline Region 3: 1.916 & High \newline(86 explicit SR expressions) & Across 86 SR expressions, \newline 40 to 170 total nodes & Lightweight and interpretable; suitable for system-level planning, propagation calculators. \\
\midrule
\textbf{SR + DE for Rain Attenuation \cite{Matondo2025RainSR}} & Hybrid SR combined with Differential Evolution model for rain-induced signal loss in 5G mmWave links & Synthetic 79000 sample dataset (3GPP RMa/UMa/UMi, 26–60 GHz) & $R^{2}=0.999$ \newline $MSE =0.026$ (urban) / $0.50$ (rural opt.) \newline $MAE < 0.04$ dB post-optimization & High (explicit analytical attenuation equations) & 60 total nodes & White-box formulas achieving $\sim 60\%$ error-reduction vs ITU-R and Crane models; suited for 5G mmWave planning tools. \\
\midrule
\textbf{Deep SR and KANs for Path Loss \cite{Anaqreh2025PathLossSR}} & Compares Deep SR and Kolmogorov–Arnold Networks for approximating standard ABG and CI path-loss models & 1000 synthetic samples (generated using ABG/CI PL models across 2--73\,GHz and 1--500\,m distances) & $R^{2} = 0.98$ (ABG) / 0.99 (CI) \newline $MSE = 0.001 / 0.0002$ \newline $MAE < 0.02$\,dB & High \newline(white-box for DSR/KAN) & Low (closed-form evaluation) & KANs outperform classical SR and GP with $\sim 99\%$ fit accuracy and interpretable mapping between inputs and path loss. \\
\midrule
\textbf{SABER for AoA and Beam Pattern Estimation \cite{Chou2025SABER}} & \raggedright SR framework for learning closed form Angle-of-Arrival and beam-pattern relations from path-loss coefficients & Measured S-parameter dataset. Stage I (26--31 GHz, $0\text{--}120^\circ$ AoA sweep, $\approx 12000$ samples, anechoic) and Stage II (28--30 GHz, RIS-aided indoor link, $\approx 8000$ samples) & $MAE = 0.396^\circ$ in free-space and $6.53 \times 10^{-7}$$^\circ$ for RIS & High (closed-form symbolic models) & LOS - 20 nodes \newline RIS - 47 nodes & Achieves sub-degree accuracy and near-optimal Root MSE (RMSE) while remaining fully interpretable; bridges classical array models and ML estimators. \\
\midrule
\textbf{SR-Aided Multi-Link Prediction \cite{pendyala2024multi}} & LTC + LSTM based mmWave SNR predictor; SR yields interpretable SNR dynamics. & NYUSIM dataset (28\,GHz, 400\,MHz BW, 64\,BS antennas, 8\,UEs, urban microcell indoor, 20\,ms sampling) & LTC achieved 0.25 dB RMSE (vs. 3.44 dB for LSTM) \newline SR model attained $R^{2}=0.825$ & High (analytical SR expression for SNR) & \textbf{116} total nodes & Interpretable SR model; LTC attained 13 x lower RMSE than LSTM, showing the accuracy–explainability trade-off. \\
\bottomrule
\end{tabular}
}
\end{table*}


\subsection{Feature Importance and Interpretability}
A distinct advantage of SR is its ability to directly provide insights into the underlying physical mechanisms. The relative frequency with which each feature appears in the final 158-node expression is presented in Table \ref{tab:feature_importance}.

\begin{table}[!b]
\centering
\caption{Relative Feature Frequency in Symbolic Expression}
\label{tab:feature_importance}
\begin{tabular}{lll}
\toprule
\textbf{Feature} & \textbf{Usage Frequency (\%)} & \textbf{Rank} \\
\midrule
$\text{SNR}_{\text{TB\_dB}}$ & $\mathbf{38.16}$ & 1 (Dominant) \\
$\text{v}_{\text{rel\_kmph}}$ & $13.16$ & 2 (Mobility) \\
$\text{N}_{\text{DMRS}}$ & $11.84$ & 3 (Channel Estimation) \\
$\text{MCS}_{\text{Modulation\_Index}}$ & $11.84$ & 4 (Spectral Efficiency) \\
$\text{Flag}_{\text{NLOS}}$ & $10.53$ & 5 (Environment) \\
$\text{MCS}_{\text{Code\_Rate}}$ & $10.53$ & 6 (Coding Gain) \\
$\text{N}_{\text{sub}}$ & $2.63$ & 7 \\
$\text{Flag}_{\text{Urban}}$ & $1.32$ & 8 \\
\bottomrule
\end{tabular}
\end{table}

The analysis of feature usage robustly confirms that the derived expression is deeply physics-grounded. It correctly identifies $\mathbf{SNR_{\text{TB\_dB}}}$ as the overwhelmingly dominant factor ($\mathbf{38.16\%}$), aligning perfectly with fundamental wireless communication theory. Following $\mathbf{SNR}$, $\mathbf{v_{\text{rel\_kmph}}}$ (relative velocity) is ranked second ($\mathbf{13.16\%}$), validating that the model successfully captures the detrimental impact of high mobility and associated Doppler spread on $\text{V2X}$ link performance.

\subsection{Feature Interaction and Sensitivity Analysis}
\label{sec:sensitivity}
The high complexity of the derived $\mathbf{158}$-node $\text{V2X}$ $\text{BLER}$ expression is a direct mathematical manifestation of the underlying physics being modeled.

\textbf{A. Primary Feature Interaction:}
The most frequently used features, $\mathbf{SNR_{\text{TB\_dB}}}$ ($\mathbf{38.16\%}$) and $\mathbf{v_{\text{rel\_kmph}}}$ ($\mathbf{13.16\%}$), are frequently co-occur within multiplicative, ratio, and power functions. This structural coupling confirms that the effect of $\text{SNR}$ on $\text{BLER}$ is non-linearly scaled and modified by the mobility factor ($\mathbf{v_{\text{rel}}}$), accurately capturing the destructive interaction between instantaneous signal strength and Doppler-induced channel time-variation.

\textbf{B. Sensitivity Analysis:}
A sensitivity analysis, calculating the partial derivative $\mathbf{\text{Sensitivity to } X = \frac{\partial Y}{\partial X}}$, was performed to gain a quantitative understanding of each feature's true marginal impact. $\mathbf{SNR_{\text{TB\_dB}}}$ consistently exhibits the largest and most stable magnitude of sensitivity. In contrast, $\mathbf{v_{\text{rel\_kmph}}}$ displays a highly non-linear and fluctuating sensitivity profile, indicating that the impact of relative velocity is intricately dependent on the specific operating point and the values of other interacting parameters. This exceptional ability to analytically confirm the model's fidelity to known channel physics is a key contribution of this work.


\subsection{Physical Validation of Model Fidelity}
Beyond overall accuracy, a critical aspect of validating the SR model is its ability to accurately reflect known physical phenomena.

\subsubsection{Physical Validation (Waterfall Curves)}
\begin{figure}[!t]
    \centering
    \includegraphics[width=\columnwidth]{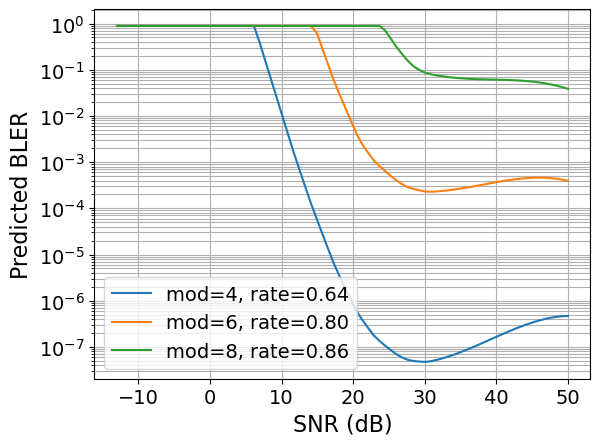}
    \caption{Predicted BLER vs. SNR (Waterfall Curves) for fixed $v_{\text{rel}}$ and different $\text{MCS}$ modulation indices. The correct shift and smooth decay validate the physical fidelity.}
    \label{fig:waterfall_curves}
\end{figure}

To provide a conclusive validation of the physical plausibility, we used the analytical formula to generate traditional $\text{BLER}$ vs. $\text{SNR}$ "waterfall curves" (Figure \ref{fig:waterfall_curves}).
\begin{itemize}
    \item \textbf{Smooth and Monotonic Decay:} The predicted $\text{BLER}$ values decrease smoothly and monotonically as the $\text{SNR}$ increases.
    \item \textbf{Physically Consistent Shifts:} Curves corresponding to higher modulation orders (e.g., $\text{Modulation Index}=8$) are correctly shifted to the right (i.e., require a higher $\text{SNR}$) compared to curves for lower modulation orders, accurately reflecting the trade-off between spectral efficiency and robustness.
\end{itemize}

\subsubsection{Physical Validation of Mobility Impact}
\label{sec:mobility_validation}
\begin{figure}[!b]
    \centering
    \includegraphics[width=\columnwidth]{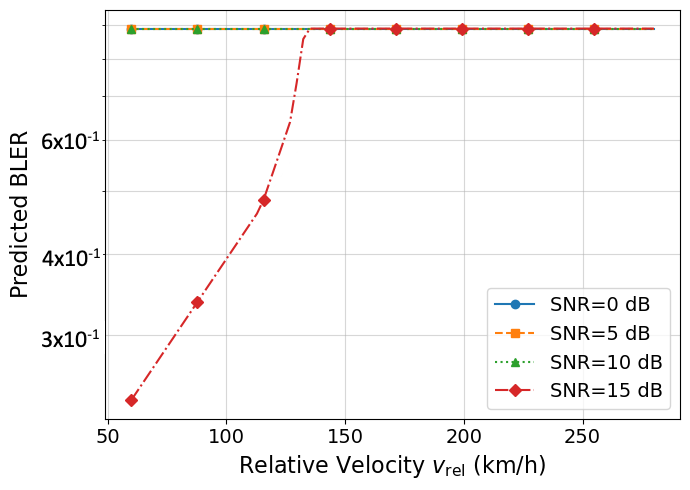} 
    \caption{Predicted BLER as a function of Relative Velocity ($v_{\mathrm{rel}}$) for different SNR levels (Modulation Index=6, Code Rate=0.8). The plot shows BLER saturation at low SNRs and severe, sharp degradation due to mobility (Doppler spread) at high SNRs.}
    \label{fig:mobility_impact_velocity}
\end{figure}

Figure \ref{fig:mobility_impact_velocity} illustrates the model's predictions for $\text{BLER}$ as a function of relative velocity ($v_{\mathrm{rel}}$). The plot clearly demonstrates:
\begin{enumerate}
    \item \textbf{MCS Operational Limits:} At low $\text{SNR}$ levels (0 dB, 5 dB, and 10 dB), the $\text{BLER}$ is predicted to be near unity ($\mathbf{10^0}$) even at the lowest velocities, confirming that the high-efficiency $\text{MCS}$ is fundamentally unsuitable.
    \item \textbf{Mobility-Induced Degradation:} For high $\text{SNR}$ (15 dB, red curve), the $\text{BLER}$ shows a sharp, non-linear increase with velocity. This degradation is a direct consequence of Doppler spread. The sharp transition from an acceptable $\text{BLER}$ ($\sim 10^{-2}$) to near-failure ($\sim 10^0$) confirms that the $\text{SR}$ model successfully captures the physics of the mobility-induced error floor.
\end{enumerate}

\subsection{Model Accuracy and Residual Analysis}
\subsubsection{Validation of Predicted vs. True BLER}
\begin{figure}[!t]
    \centering
    \includegraphics[width=\columnwidth]{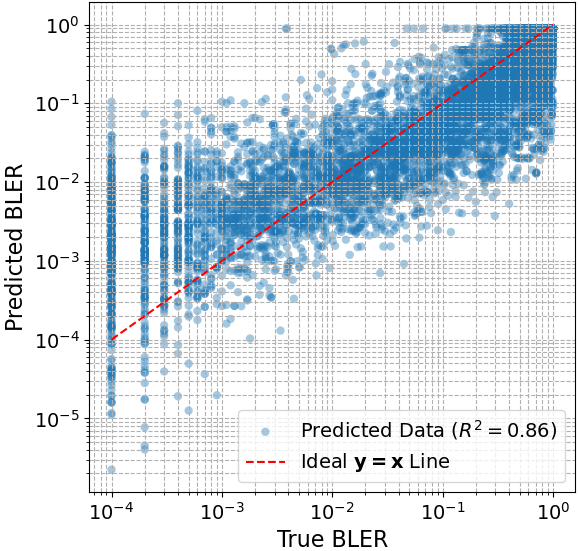}
    \caption{Predicted BLER vs. True BLER (Log-Log Scale) on the test set. The tight cluster around the $y=x$ line confirms generalization.}
    \label{fig:scatter_bler}
\end{figure}

A crucial step is the visual comparison of its predictions against the ground truth. Figure \ref{fig:scatter_bler} presents a scatter plot comparing the predicted $\text{BLER}_{\text{pred}}$ values from our $\text{SR}$ model against the true $\text{BLER}_{\text{true}}$ values on the unseen test set. The dense clustering of data points tightly around the $\mathbf{y=x}$ line, extending from very low $\text{BLER}$ values ($\approx 10^{-4}$) up to $1$, provides strong evidence of the model’s high predictive accuracy and robust generalization capability.

\subsubsection{Residual Analysis}
\begin{figure}[!t]
    \centering
    \includegraphics[width=\columnwidth]{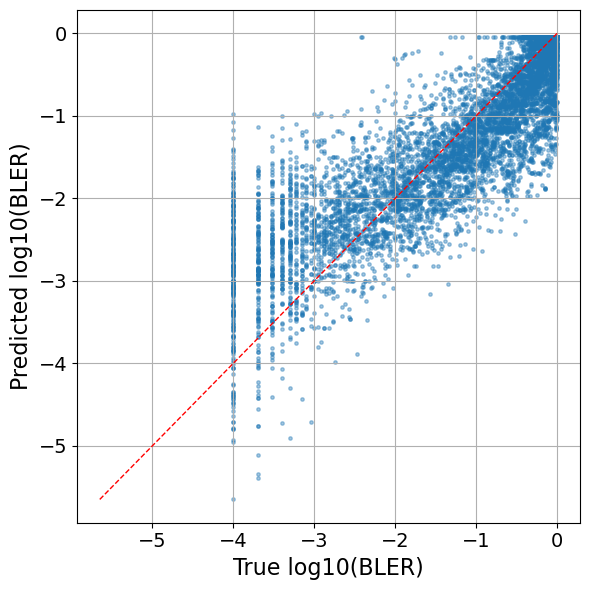}
    \caption{Predicted $\log_{10}(\text{BLER})$ vs. True $\log_{10}(\text{BLER})$. The narrow, centered band confirms consistent and unbiased error distribution.}
    \label{fig:residual_log}
\end{figure}

Detailed analysis of the residuals, computed on the $\log_{10}(\text{BLER})$ scale, further validates the high quality and unbiased nature of our model:
\begin{itemize}
    \item Mean Residual: $\mathbf{-3.472 \times 10^{-3}}$ ($\approx 0$) - Confirms the model is \textbf{unbiased}.
    \item Standard Deviation of Residuals: $\mathbf{0.3922}$ - Indicates that the prediction errors are tightly bounded and consistently small.
\end{itemize}
Figure \ref{fig:residual_log} graphically illustrates the distribution of these residuals, showing a narrow, centered band without discernible patterns, which provides additional confidence in the reliability and consistency of the derived analytical $\text{BLER}$ formula.

\subsection{Summary of Advantage}
In summary, the proposed $\text{SSRV-BLER}$ model achieves \textbf{near–deep-learning accuracy} while maintaining full analytical transparency and sub-millisecond inference cost ($\mathbf{\sim 158}$ operations). By providing an instantly computable, closed-form mapping between $\text{BLER}$, $\text{SNR}$, $\text{MCS}$, and mobility, it is uniquely suitable for real-time $\text{V2X}$ link adaptation and system-level simulation where both \textbf{accuracy and computational efficiency are critical.}

\section{Conclusion}
We have presented a SSRV framework for accurate and interpretable prediction of BLER in V2X wireless networks. The model captures complex, nonlinear interactions among system parameters, including SNR, mobility, modulation, coding, and environmental factors. With only 158 computational nodes, the derived symbolic expression enables ultra-fast inference suitable for real-time embedded deployment, overcoming the computational overhead of DNNs. Comparative analysis demonstrates that SSRV outperforms fixed-form regressions and achieves accuracy close to black-box neural models while maintaining full interpretability. Residual and feature sensitivity analyses validate the model’s physical consistency and highlight dominant contributors to BLER, such as SNR and relative velocity. Waterfall curves and mobility-dependent BLER trends predicted by the model closely match expected physical behavior, confirming its utility for scenario evaluation and network design. Overall, the SSRV approach provides a unified framework combining predictive accuracy, analytical transparency, and computational efficiency, making it a practical and insightful tool for V2X system optimization, adaptive scheduling, and real-time performance monitoring.

\section*{Appendix: Full Discovered Expression}
\lstset{
    basicstyle=\fontsize{5.5}{6.2}\ttfamily,   
    breaklines=true,                        
    breakatwhitespace=true,                    
    columns=fullflexible,                   
    frame=none,                                
    numbers=none,                             
    tabsize=2,                                
    showstringspaces=false,                    
    keepspaces=true,                        
    xleftmargin=0em, xrightmargin=0em,       
    aboveskip=0.4em, belowskip=0.4em,        
    postbreak=\mbox{},                       
}

The final symbolic expression for $Y = -\ln(\text{BLER})$ is a 158-node symbolic tree.
All variables are standardized versions of the input features:

\begin{itemize}
    \item $X_0$: $\mathbf{\text{SNR}_{\text{TB\_dB}}}$
    \item $X_1$: $\mathbf{\text{MCS}_{\text{Code\_Rate}}}$
    \item $X_2$: $\mathbf{\text{MCS}_{\text{Modulation\_Index}}}$
    \item $X_3$: $\mathbf{v_{\text{rel\_kmph}}}$
    \item $X_4$: $\mathbf{N_{\text{sub}}}$
    \item $X_5$: $\mathbf{N_{\text{DMRS}}}$
    \item $X_6$: $\mathbf{\text{Flag}_{\text{Urban}}}$
    \item $X_7$: $\mathbf{\text{Flag}_{\text{NLOS}}}$
\end{itemize}

\begin{lstlisting}[caption={Full 158-node SR Expression for $Y = -\ln(\text{BLER})$}, label={lst:full_expression}]
max(0.117, add(add(mul(cos(SNR_TB_dB), sub(SNR_TB_dB, MCS_Modulation_Index)), mul(MCS_Modulation_Index, -0.781)), sub(add(add(add(min(SNR_TB_dB, add(neg(sin(v_rel_kmph)), Flag_NLOS)), sub(SNR_TB_dB, MCS_Modulation_Index)), sub(add(add(add(add(min(SNR_TB_dB, add(cos(SNR_TB_dB), neg(min(sqrt(SNR_TB_dB), v_rel_kmph)))), mul(MCS_Modulation_Index, -0.781)), sub(add(add(sub(sub(add(min(sub(SNR_TB_dB, MCS_Modulation_Index), cos(SNR_TB_dB)), min(SNR_TB_dB, sqrt(v_rel_kmph))), sub(MCS_Code_Rate, SNR_TB_dB)), sub(MCS_Modulation_Index, min(SNR_TB_dB, add(cos(N_DMRS), neg(sin(sin(N_sub))))))), min(SNR_TB_dB, add(N_DMRS, add(neg(v_rel_kmph), cos(N_DMRS))))), min(SNR_TB_dB, add(neg(sin(sin(v_rel_kmph))), cos(N_DMRS)))), sub(MCS_Code_Rate, SNR_TB_dB))), min(sin(sub(SNR_TB_dB, Flag_NLOS)), add(neg(sub(div(MCS_Modulation_Index, sqrt(SNR_TB_dB)), neg(max(sin(v_rel_kmph), MCS_Code_Rate)))), neg(min(pow(-0.548, N_DMRS), v_rel_kmph))))), min(SNR_TB_dB, add(neg(min(sqrt(SNR_TB_dB), v_rel_kmph)), neg(min(pow(-0.548, N_DMRS), v_rel_kmph))))), sub(MCS_Code_Rate, SNR_TB_dB))), min(SNR_TB_dB, add(min(SNR_TB_dB, add(add(neg(SNR_TB_dB), cos(SNR_TB_dB)), neg(sin(v_rel_kmph)))), add(N_DMRS, neg(sin(v_rel_kmph)))))), sub(MCS_Code_Rate, SNR_TB_dB))))
\end{lstlisting}

\section*{Acknowledgments}
The authors used AI tools solely for editing and improving the clarity and grammar of the text. All technical content, research findings, and scientific interpretations presented in this paper were generated exclusively by the authors.

\bibliographystyle{IEEEtran}
\bibliography{main}

\vfill

\end{document}